\documentclass[prl, preprintnumbers, amsmath, amssymb, superscriptaddress, twocolumn]{revtex4}
\usepackage{graphicx, color, dcolumn, bm, natbib, amssymb, amsmath}
\usepackage[final=true]{hyperref}
\usepackage[utf8x]{inputenc}
\usepackage[russian]{babel}
\usepackage[normalem]{ulem}
\usepackage{hyphenat}

\begin{document}

\title{Stability boundaries of the skyrmion phase in non-centrosymmetric ferromagnets with Dzyaloshinskii-Moriya interaction}

\author{S. V. Grigoriev}
	\affiliation{NRC ``Kurchatov Institute'', Petersburg Nuclear Physics Institute, Gatchina
188300, Russia}
	\affiliation{St.Petersburg State University, 7/9 Universitetskaya nab., 199034
St.~Petersburg, Russia}	
	
	\author{V.E. Timofeev}
\email{viktor.timofeev@spbu.ru}
	\affiliation{NRC ``Kurchatov Institute'', Petersburg Nuclear Physics Institute, Gatchina
188300, Russia}
	\affiliation{St.Petersburg State University, 7/9 Universitetskaya nab., 199034
St.~Petersburg, Russia}	
    \affiliation{Ioffe Institute, St.~Petersburg, 194021, Russia}		
	
	\author{D.N. Aristov}
	\affiliation{NRC ``Kurchatov Institute'', Petersburg Nuclear Physics Institute, Gatchina
188300, Russia}
	\affiliation{St.Petersburg State University, 7/9 Universitetskaya nab., 199034
St.~Petersburg, Russia}	 
    \affiliation{Ioffe Institute, St.~Petersburg, 194021, Russia}	

\begin{abstract}
Stability boundaries of the skyrmion lattice in non-centrosymmetric bulk ferromagnets with the Dzyaloshinskii-Moriya interaction in external magnetic field are discussed. We compare the classical energies of the spin configuration of the conical helix and skyrmion lattice within the framework of the stereographic projection approach. It is well known that at low temperatures the skyrmion lattice loses energetically
 to the conical helix in the entire range of fields, $0 < H < H_{c2}$, where $H_{c2}$ is the transition field to the polarized collinear phase,
and $g\mu_B H_{c2} \ll T_c$. We show that taking into account the dipole interaction does not qualitatively change the situation. However, the possibility of fluctuations in the absolute value of the equilibrium local magnetization in the Ginzburg-Landau functional leads, with increasing temperature, $T$, to the skyrmion lattice becoming energetically more favorable than the conical helix in a certain range of fields. We show that it occurs already in the first order of small parameter, $\propto g\mu_B H_{c2}/|T-T_c|$, at the level of mean field theory. 
\end{abstract}

\maketitle

{\it Introduction.} The appearance of a skyrmion lattice in a small region of the $(H-T)$ magnetic phase diagram of manganese monosilicide MnSi and its related compounds continues to attract  interest and to stimulate a large number of both experimental and theoretical studies. The region of stability of the skyrmion phase {for bulk samples} is characterized by a modest place near the critical temperature $T_c$ in a certain range of fields, but it was discovered in {\it all without exception} magnetic compounds with a $B20$ type structure \cite{Ishimoto1, Ishimoto2, Lebech_95,Grigoriev2006, Grigoriev2007_1, Grigoriev2007_2, MuhlbauerSience09_v323,Adams_PRL_2011, Grigoriev_Jetp_Lett_14, PfleidererJPCM10_v22, NeubauerPRL09_v102,Muenzer_PRB, Moskvin_PRL_2013, Adams_PRL_2012, White_PRL_2012, Tokunaga}. The nature of the phenomenon was established in early neutron scattering experiments as a static magnetic modulation propagating perpendicular to the applied field in  MnSi \cite{Lebech_95, Grigoriev2006} and solid solutions Fe$_x$Co$_{1-x}$Si  \cite{Ishimoto1, Ishimoto2, Grigoriev2007_1, Grigoriev2007_2}. Later, the complete picture of the structure was established, which is a two-dimensional hexagonal magnetic lattice with ${\bf k}_{h(1,2,3)} \perp \bf{H}$ in MnSi \cite{MuhlbauerSience09_v323} and other transition metal monosilicide compounds Mn/FeSi, Mn/CoSi, Fe/CoSi \cite{PfleidererJPCM10_v22, NeubauerPRL09_v102, Muenzer_PRB}, FeGe \cite{Moskvin_PRL_2013}, Cu$_2$OSeO$_3$ \cite{Adams_PRL_2012, White_PRL_2012} and $\beta$-Mn-type Co-Zn-Mn \cite{Tokunaga} alloys.

It is well known that the magnetic structure of MnSi is built on a hierarchy of interactions: ferromagnetic exchange interaction, antisymmetric Dzyaloshinskii-Moriya (DM) interaction, as well as anisotropic exchange interaction and cubic anisotropy \cite{Bak, Kataoka_1}. A magnetic system based on these interactions in zero field is ordered into a magnetic helix with a wave vector whose magnitude is determined by the ratio between the ferromagnetic and DM interactions, $k_0 = D/J$. With the application of a magnetic field, the spiral firstly transforms into a single-domain conical structure with ${\bf k} \parallel {\bf H}$. Then the conical structure transforms into a collinear, fully polarized structure at a critical field $H_{c2}$. The conical phase is observed in the entire temperature range below the critical temperature $T_c$. 
However, in a narrow range of fields near $T_c$, competing with the conical structure, a skyrmion lattice (SL) \cite{MuhlbauerSience09_v323, Adams_PRL_2011, Grigoriev_Jetp_Lett_14} arises, which is also observed in numerical simulations \cite{Buhrandt2013}.
The authors of \cite{MuhlbauerSience09_v323} proposed the Landau-Ginzburg mean-field model describing a single-domain multi(triple)-${\bf k}$ structure, which they called a skyrmion lattice. It was argued that SL is less energetically favorable both compared to a simple helical structure in zero field and compared to a conical structure with an applied magnetic field \cite{Butenko2010}. It was suggested that Gaussian thermal fluctuations near $T_c$ stabilize the triple-${\bf k}$ state at some field $H_A \approx 0.4\, H_{c2}$ \cite{MuhlbauerSience09_v323}. It was also noted that at a field of $H_A \approx 0.4\, H_{c2}$, a configuration with the highest skyrmion density is realized \cite{Rybakov,Wilson}.
As an alternative to the three-$\mathbf{k}$ structure approach, it is worth mentioning the analysis of the Ginzburg-Landau functional with a soft modulus, when the skyrmion configuration is modeled on a disk, and the variational procedure leads to a system of equations for the magnetization rotation angle and the absolute value of the magnetization \cite{Rossler2010,Laliena2018}. An analysis of the interaction energy and the role of fluctuations above the magnetic ordering point was done in  \cite{Janoschek2013}.

The issue of stability of the skyrmion lattice is actively discussed in connection with experiments performed with thin films of B20 compounds \cite{Yu_2010, Yu_2011, Tokura_2012}. Lorentz transmission electron microscopy (TEM) shows that the stability of the SL critically depends on the film thickness: the thinner the film, the higher is the stability and the more extensive the SL region becomes on the magnetic phase diagram (H-T) of the studied compounds \cite{Yu2015}. Theoretical and experimental work shows that in order to ensure the stability of the SL, the system must either have uniaxial anisotropy, reducing the symmetry of the problem to two-dimensionality, or must have a surface/interface, providing a restricted geometry \cite{Rybakov, Wilson, Monchesky, Kiselev1, Du_Nano_Lett}. Note that for films in the usual experimental setup, the field is directed perpendicular to the surface, and the conical structure cannot be realized when the pitch of the magnetic helix becomes smaller than the film thickness. Instead, a helix with a vector ${\bf k}$ in the plane is realized, which loses in energy to the skyrmion lattice \cite{Timofeev2021,Timofeev2022}. It is interesting to note that the boundaries of the skyrmion phase not only in the field but also in temperature significantly depend on the shape of the sample, that is, on the demagnetizing factor \cite{bauer_PRB}.

In this paper we discuss the competition between the skyrmion lattice and the conical structure in a magnetic field and describe the stability boundaries of the skyrmion phase in the magnetic phase diagram of non-centrosymmetric ferromagnets with the Dzyaloshinskii-Moriya interaction. First, we discuss the case of low temperatures, $T\ll T_{c}$, when the conical spiral is slightly more favorable in energy than the skyrmion lattice in the entire range of fields from $0$ to $H_{c2}$. However, the energy difference in a certain field, $H_{A}$, turns out to be minimal, about $0.7\%$ of the characteristic energy that distinguishes the spiral phase from the fully polarized phase, $Ak_{0}^{2} = g\mu_{B} H_{c2}$. This field, $H_{A} \approx 0.4\, H_{c2}$, is characterized by the equality of the average magnetization of the conical phase and the skyrmion lattice.

Further consideration of the dipole-dipole interaction reduces the energy difference, but does not change the situation qualitatively. The possibility of fluctuations in the absolute value of the local magnetization as the temperature approaches $ T_{c}$ leads to a gain in energy for the skyrmion phase in a certain range of fields. It turns out that the skyrmion lattice becomes energetically more favorable than the conical phase already in the first order of smallness in the parameter $ J k_0^2/|T-T_c|$, at the  mean field level of theory. We find an estimate for the stability region of the skyrmion phase in terms of the model parameters.

{\it Model: exchange + DM interaction.}
We consider the energy density of a magnet of the following form
\begin{equation}
{\cal H}= \tfrac12  J (\nabla \mathbf{m} )^2 + D\,   \mathbf{m} \cdot \nabla\times \mathbf{m}  -  g\mu_B \mathbf{m}\cdot\mathbf{H}
+ H_{d} \,. 
\label{Ham}
\end{equation}
At low temperatures, the local magnetization $\mathbf{M} = g\mu_B \mathbf{m}$ saturates to $|\mathbf{M}| = g\mu_B S /v_0 $, where $S$ is the magnitude of the localized moment, $v_0$ is the volume of the unit cell. The field $\mathbf{H}$ is directed along the $\hat z$ axis. The form of the dipole-dipole interaction energy $H_{d}$ is not written out explicitly; its role in the case of cubic magnets is reduced mainly to the demagnetization field, see below.

Minimization of the classical ground state energy of the system (taking into account ferromagnetic exchange, DM interaction, anisotropic exchange, interaction with the magnetic field and dipole energy) for a simple helix and a conical helix phase is considered in \cite{Maleyev06PRB}, and for a skyrmion lattice - in \cite{MuhlbauerSience09_v323}, see also \cite{Timofeev2021}. It is known\cite{Kataoka_1, Bak} that the minimum of the ground state energy is realized at the wave vector of the structure \[ k_0 = D/J \ll 1 \,,\] which is typical for both the conical structure and the skyrmion lattice. This is confirmed by the entire set of neutron scattering experiments \cite{Lebech_95, Grigoriev2006, MuhlbauerSience09_v323, Grigoriev_Jetp_Lett_14, Chubova_JETP_2017}.

Let us consider the ordering in the phase of a conical helix, which is described by the vector $\mathbf{n} = \mathbf{m} /|\mathbf{m}|$ of the following form
\begin{equation}
 \mathbf{n} =  (\cos\alpha \cos k_0 z,\cos\alpha \sin k_0 z, \sin\alpha)\,,
 \label{n-con}
\end{equation}
where the cone angle is $(\pi/2 - \alpha)$. Varying the expression for the energy \eqref{Ham}, we find $k_0 = D/J$, and the energy at the minimum is given by \cite{Maleyev06PRB}
\begin{equation}
E_{con} = -\tfrac12 {m Ak_0^2} \cos^2\alpha - g\mu_B m H \sin\alpha\,,
\label{eq1}
\end{equation}
where $A= Jm$ and $m=|\mathbf{m}|$.
Minimizing the expression \eqref{eq1} over the angle $\alpha$, we find that
\begin{equation}
\label{eq2}
\sin\alpha =  {H}/{H_{c2}} \,, \quad H_{c2} =  Ak_0^2/g \mu_B    \,. 
\end{equation}
The value $ \alpha = \pi/2$ is achieved at $H=H_{c2}$. For the archetypal compound MnSi, it was shown in neutron experiments that the relation \eqref{eq2} is well satisfied over the entire temperature range up to $T_c$ \cite{Kizhe}.

As a result, the classical energy of the conical structure is equal to:
\begin{equation}
\label{eq3}
E_{con} =  - \frac {mAk_{0}^2}2 \left [  1 + \frac{  H^2}{H_{c2}^{2}} \right ] 
= - \frac {g \mu_B m  }2 \left[ H_{c2} + \frac{  H^2}{H_{c2} } \right]
\,.
\end{equation}
from here it is clear that $H_{c2} $ corresponds at $H=0$ to the difference in energies of the ferromagnetic state and the Dzyaloshinskii helix \cite{Kataoka_1, Maleyev06PRB}. Note also that the uniform component of magnetization along the field is given by the expression $g \mu_B m_{3} = - d E/dH $ and depends linearly on the field, while the susceptibility, $\chi = - d^{2}E/dH^{2}$, in the conical phase is equal to a constant, $\chi_{c} = g \mu_B m /H_{c2} = (g \mu_B)^{2} J/D^{2}$.

The case of a skyrmion structure can be described in the stereographic projection method. A triangular skyrmion lattice is given with good accuracy by the sum of radially symmetric profiles of individual skyrmions, centered at the lattice sites \cite{Timofeev2021,Timofeev2022}. Each individual profile corresponds to the expression
\begin{equation}
 \mathbf{n} =  (-\sin\varphi  \sin \theta(r) ,\cos\varphi \sin \theta(r) , -\cos \theta(r)  )\,,
 \label{n-skyr}
\end{equation}
where $\mathbf{r} = r (\cos\varphi, \sin\varphi )$. The energy-optimizing function $\theta(r)$ depends on the magnetic field strength. At the same time, the calculated optimal distance, $2r_{0}$, between skyrmions \cite{Timofeev2021} depends much less on the field in the region of interest, $0.4< H/H_{c2}<0.6$, and is given, with good accuracy, by the ratio $J/D$, so that $r_{0} = \pi/k_{0}$.

\begin{figure}[h]
	\centering
 	\includegraphics[width=0.9\columnwidth]{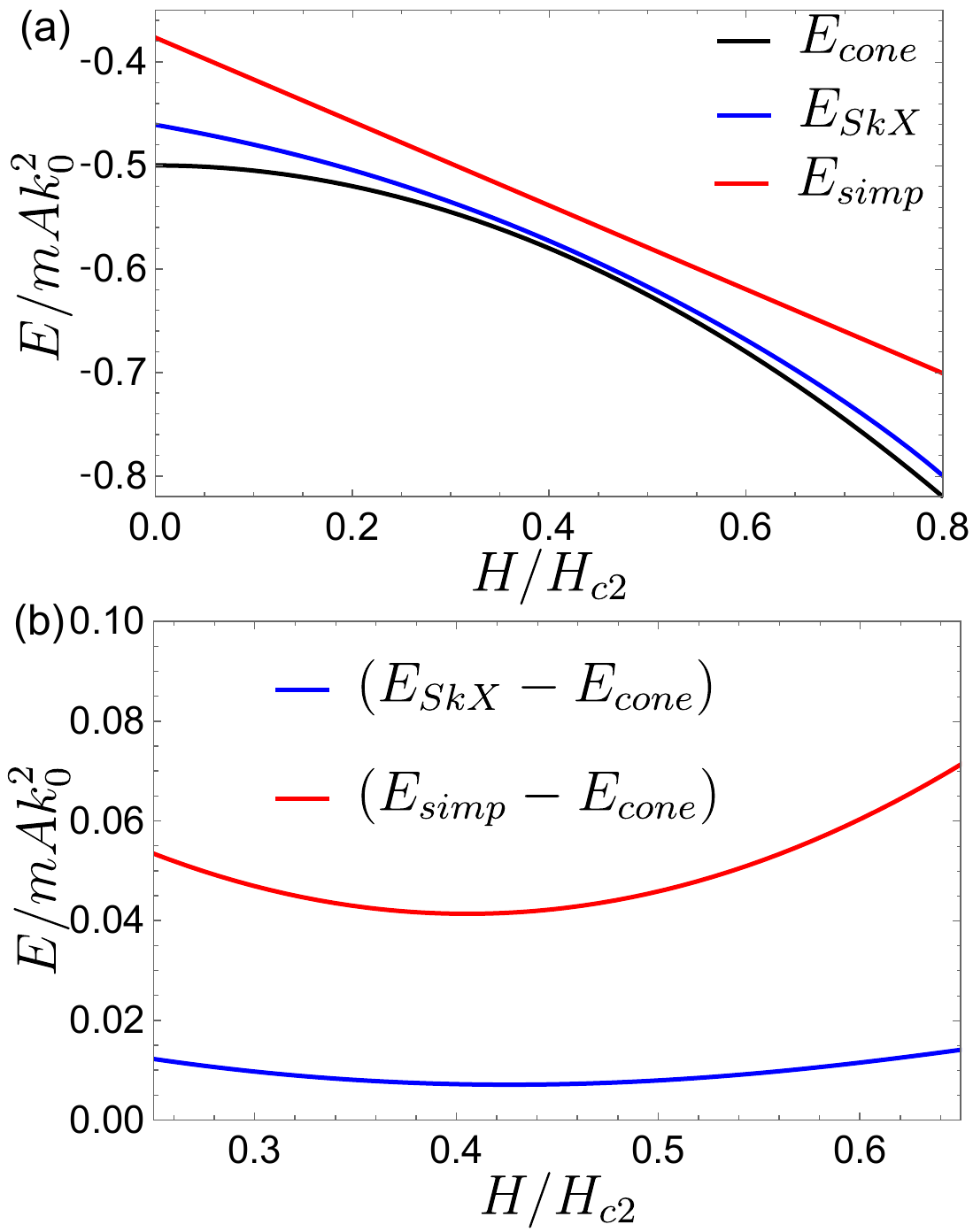}\\
	\caption{ The dependence of the interaction energy on the magnetic field for the conical structure, Eq.\eqref{eq3}, and the skyrmion lattice, Eq.\eqref{eq7} and \eqref{endiff}. The field $H$ is normalized by the critical field, $H_{c2}$, and the energy $E$ is normalized by $mAk_{0}^2$.} 
	\label{fig1}
\end{figure}

A calculation of the skyrmion lattice energy, $E_{sk}$, using the stereographic projection method shows that $E_{sk}$ is always slightly higher than the conical helix energy, $E_{con}$. These energies are almost identical at the field value $H_{*} \simeq 0.43 H_{c2}$, with the minimum difference
\begin{equation}
\min (E_{sk}  - E_{con})  \simeq   0.007 \,  {mAk_{0}^2} \,,
\label{estimEnSkCon1}
\end{equation}
weakly depending on $T$ over a wide temperature range. Let us try to give a qualitative explanation for it, by considering a simplified model in which the dependence $\theta(r)$ is given by a linear law, $\theta(r) = k_{0} r$, so that the direction of magnetization coincides with the direction of the field at the boundary between skyrmions, $r_{0}=\pi/k_{0}$.

By analogy with Eq.\eqref{eq1}, we represent the dependence of the classical energy of the skyrmion lattice on the magnetic field in the form of two terms, a bilinear contribution in spins and a Zeeman contribution:
\begin{equation}
\label{eq4}
E_{sk} =  -\tfrac 12 {mAk_{0}^2}    \langle \tilde n_{\perp}^{2}\rangle -  g\mu_B m  H  \langle n_{\|}\rangle \,,
\end{equation}
where $n_{\|} = -\cos\theta$, $\tilde n_{\perp}^{2} = 1 -\sin^{2} (\theta)/\theta^{2}$, and $\langle \ldots \rangle$ means averaging over a circle of radius $r_{0}$.

The averages arising in Eq.\ \eqref{eq4} are easily calculated:
\begin{equation}
\langle n_{\|}\rangle = 4/ \pi^{2} \,, \quad  \langle \tilde n_{\perp}^{2}\rangle \simeq 0.753 \,.
\label{eq5}
\end{equation}
Then the classical energy of the skyrmion lattice in a magnetic field is:
\begin{equation}
\label{eq7}
E_{sk} =  -\tfrac 12 {mAk_{0}^2} \left( 0.753 + \frac{8}{\pi^2}\frac{  H}{H_{c2} } \right) \,.
\end{equation}

The condition $d(E_{sk} -E_{con})/dH = 0$ determines the value of the field, $H_{A}$, at which the difference between the two energies, the conical structure, Eq.\eqref{eq3}, and the skyrmion lattice, Eq.\eqref{eq4}, is minimal. We find
\begin{equation}
\label{eq9}
H_{A} = \frac{4}{\pi^2} H_{c2} \simeq 0.405 H_{c2} \,,
\end{equation}
at the same time
\begin{equation}
\left. (E_{sk}  - E_{con})\right|_{H=H_{A} } \simeq   0.041 \,  {mAk_{0}^2} 
\label{endiff}
\end{equation}

We see that in the simplified model, $\theta(r) = k_{0} r$, the energy difference is slightly larger than in the more accurate calculation. This clarifies the value of the field $H_{A}$, which has a simple geometric nature independent of the details of the Hamiltonian. Note that according to the definition of magnetization, $H_{A}$ is the field at which the field-directed average magnetic moment of the skyrmion lattice and the conical structure are equal.
\footnote{In \cite{Wilson} it was noted that the skyrmion density has maximum at this field value. Such a coincidence, however, is violated when the anisotropy is taken into account, see Fig. 2 in \cite{Wilson}.}

Experimentally, in bulk samples, the skyrmion lattice is observed only near $T_c$ and in a certain range of fields, and not at a single point $H_A$. One has to find a reason for the small gain in the energy of the skyrmion lattice at high temperatures. Note that contributions to the classical energy of the second order (cubic anisotropy in the case of MnSi) can only slightly affect the stability boundaries of the conical phase and the skyrmion lattice. Let us analyze first the role of the dipole-dipole interaction, which is weak but long-range. It turns out that the magnitude and the character of the demagnetization field $H_d$, caused by dipole forces, differ for the conical structure and the skyrmion lattice.

{\it Model: exchange + DM interaction + demagnetization.}
We have the general relation $H_{in} = H_{ext} - N_{3}m_{3}$, where $N_{3}$ is the demagnetization factor, depending on the shape of the sample, and $m_{3}$ is the average value of magnetization along the field, $m_{3} = \chi_{c} H_{in}$.
For a conical structure, the demagnetization field, $N_{3}m_{3}$, turns out to be directly proportional to the magnetic field. In Eqs.\  \eqref{eq1} and \eqref{eq3}, the external field must be replaced by the internal one, according to
\begin{equation}
H_{in, con} = H_{ext} /(1 + \chi_{c} N_{3}) \equiv  H_{ext}/a_1
\label{eq10} 
\end{equation}
As the experiments and the above calculations show, the susceptibility in the conical phase does not change for all fields and temperatures, $\chi_{c} = const$.
It is also shown that for cubic magnets with DM interaction, the susceptibility is less than unity, for example, $\chi_{c} \approx 0.3$ for MnSi \cite{bauer_PRB}. From this we conclude that the coefficient $a_1$ does not depend on $H_{ext}$ and $T$ and does not differ much from unity.

The situation is different for the skyrmion lattice. The skyrmion lattice structure turns out to be quite rigid in terms of the period. The average value of the magnetization along the field, $m_{3}$, is not proportional to the magnetic field, $H$. In the simplified model above, \eqref{eq5}, $m_{3} = 4m /\pi^{2}$ is independent of the magnetic field, but depends on the temperature.
Note that, despite the one-dimensional nature of individual skyrmions, the demagnetization factor for the skyrmion lattice and for the conical structure is the same, $N_{3, sk} =N_{3}$. This is because the main contribution to $N_{3, sk}$ comes from large distances, $r\gg k_{0}^{-1}$, see \cite{Grigoriev2011}. As a result, for the simplified model we have the relation for the internal field $H_{in, sk} = H_{ext} -N_{3} m_{3}$.

The energy difference \eqref{eq3}, \eqref{eq7} now looks like
\begin{equation}
\frac{ E_{sk}  - E_{con}} {mAk_{0}^2}    =    
0.123 - \frac{4}{\pi^2}\frac{ H_{in,sk}}{H_{c2} } + \frac{  H_{in,con}^2}{ 2 H_{c2}^{2}} \,.
\label{diffEn0}
\end{equation}
Substituting here the expressions for $H_{in,con}$, $H_{in,sk}$, we find that this energy difference is minimal at $H_{ext} = ({4}/{\pi^{2}}) a_{1}^{2} \,H_{c2} $. Setting in the skyrmion phase $ m_{3} = ({4}/{\pi^{2}}) m = ({4}/{\pi^{2}}) \chi_{c} H_{c2}$, we have at the minimum
\begin{equation}
  \begin{aligned}
 \frac{ E_{sk}  - E_{con}} {mAk_{0}^2}  
 & = 0.041 -  \tfrac{8}{\pi^4}\chi_{c} ^{2}N_{3}^{2} 
\label{difEnDemag0}
\end{aligned}
\end{equation}
Assuming here $ N_{3}\simeq 1/3$, $\chi_{c} \simeq 0.3$, we see that taking into account the dipole forces in the simplified model reduces the energy difference, although it does not lead to a gain in energy for the skyrmion phase.

Returning to the full calculation, in which the difference between the energies of the conical and skyrmion phases is noticeably smaller, \eqref{estimEnSkCon1}, we can write in the skyrmion phase the refined expressions for the energy and magnetization. Near the characteristic value, $H_{*} \simeq 0.43 H_{c2}$, at which the energy difference, $(E_{sk} - E_{con})$, is minimal, we have a dependence on the internal field, $H$, of the following form
\begin{equation}
 \begin{aligned}
  E_{sk}    & \simeq mAk_{0}^2 
  \left( \epsilon_{*} - \frac{H_{*}(H-H_{*})  }{ H_{c2}^{2}} -   \epsilon''_{*} \frac{(H-H_{*})^{2} }{2H_{c2}^{2}} \right)
\\ 
 m_{3}  
 & \simeq m \left( \frac{H_{*}}{H_{c2}}   + \epsilon''_{*} \frac {H-H_{*}} {H_{c2}}\right)   
 = \chi_{c} H_{*} + \chi_{sk} (H-H_{*})
\end{aligned}
\label{EnSkLin}
\end{equation}
where $\epsilon_{*} \simeq 0.60$, and the calculated value $\epsilon''_{*} = \chi_{sk}/ \chi_{c}\simeq 0.7 $ is consistent with the observed decrease in the differential susceptibility in the skyrmion phase \cite{bauer_PRB,Bauer2016}. Note that the simplified model above corresponds to $\epsilon''_{*}=0$.

From the relation $H_{in, sk} = H_{ext} -N_{3, sk} m_{3}$ we obtain in the leading order
\begin{equation}
H_{in, sk} = \frac{H_{ext}- 
N_{3} (\chi_{c} -\chi_{sk})H_{*} }{1+N_{3} \chi_{sk} }  \equiv \frac{H_{ext} }{a_{2}} - (\alpha-1) H_{*}
\,, 
\label{inextSk}
\end{equation}
here $a_{2} = 1+N_{3} \chi_{sk} $, $\alpha =a_{1}/a_{2}$. Instead of Eq. \eqref{diffEn0} we now use expressions \eqref{eq10}, \eqref{EnSkLin}, \eqref{inextSk}, and find in the minimum
\begin{equation}
 \begin{aligned}
 \frac{ E_{sk}  - E_{con}} {mAk_{0}^2}    &=    
 0.007 -    \frac{   H_{*}^2}{  2H_{c2}^{2}} 
 \frac{N_{3}^{2}\chi_{c}(\chi_{c}-\chi_{sk})}{1-N_{3}^{2} \chi_{c} \chi_{sk} } 
  \,, 
\label{diffEn1min}
\end{aligned}
\end{equation}
in this case, the correction due to demagnetization in Eq.\ \eqref{difEnDemag0} is obtained from the second term in \eqref{diffEn1min} by substituting $H_{*} = 4H_{c2}/\pi^{2}$, $\chi_{sk}=0$. For the above values of $N_{3}$, $\chi_{c}$, $ \chi_{sk} $, we find that the minimum is attained at the field value $H_{ext} \simeq 0.52 H_{c2}$. In this case, the second term in \eqref{diffEn1min} only slightly reduces the energy difference, and the conical phase still remains more favorable in energy. We also calculated the corrections due to demagnetization, going beyond the first terms of the expansion, Eq. \eqref{EnSkLin}, and the qualitative conclusion about the slight advantage of the conical phase has not changed.

{\it Accounting of fluctuations near $T_{c}$.}
We have shown above that the energy of the ground state of the skyrmion phase in a bulk (3D) sample is always slightly higher than the energy of the conical spiral under the assumption of \emph{homogeneity of the modulus} of the local magnetization. This means that at low temperatures, when the magnetization is maximum in absolute value, the skyrmion phase is unfavorable. Let us now consider what changes at high temperatures, near $T_c$, when the modulus of magnetization $m(\mathbf{r})$ in the sample can be inhomogeneous.

To do this, we represent the free energy of a magnet, ${\cal F}$, near $T_c$ as a functional integral over the magnetization vector $\mathbf{m} (\mathbf{r})$.
\begin{equation} 
\begin{aligned}
e^{-\beta{\cal F}} &= \int\! {\cal D}\mathbf{m}  \exp \left( - \beta\!\int\! d\mathbf{r} \, 
 F (\mathbf{r} ) \right)  \,, \\ 
F(\mathbf{r})&={\cal H}+ A_{2} |\mathbf{m}|^2 +  A_{4} |\mathbf{m}|^4
\end{aligned}
\end{equation}
here $\beta = 1/T$, and the coefficients $A_{2} \sim (T -T_c) \equiv T_{c} \tau$, $A_{4} \sim T$ can be found from the high-temperature expansion \cite{Utesov2017}. Near the transition, the following terms of the expansion in even powers of $\mu$ can be omitted.

Below we omit for simplicity  the term $H_{d}$ in Eq. \eqref{Ham} and write
\begin{equation}
\begin{aligned}
{\cal H} & =    J k_{0}^{2}\left(  m^{2} \, \Omega 
+ m b 
+ \tfrac 12   (\tilde {\nabla} m )^2 \right) 
 \,,    \\ 
 \Omega & =  \frac12   (\tilde {\nabla} \mathbf{n} )^2 +   \mathbf{n} \cdot \tilde {\nabla}\times \mathbf{n}  \,,  \quad 
 b   =  -\frac{ g\mu_B }{ J k_{0}^{2}} \, \mathbf{n}\cdot\mathbf{H}  \,, 
 \end{aligned}
\label{HamGL}
\end{equation}
here we use dimensionless distances and gradients, $\tilde{\mathbf{r}}= k_{0}\mathbf{r} $, $\tilde {\nabla} = k_{0}^{-1} \nabla$. Contrasting  our approach to the works \cite{Rossler2010,Laliena2018}, we neglect below the last term, $(\tilde {\nabla} m )^2$, in Eq. \eqref{HamGL}. In the temperature range of our interest, the consideration of this term yields small corrections, ${\cal O} (\epsilon_0 \kappa^2) $, to Eq. \eqref{deltaE}.

We consider the temperature range below the critical point, $T<T_c$, ($A<0$), and solve the problem iteratively. At sufficiently low temperatures, $|T- T_c| \sim |A_{2} | \gg J k_{0}^{2} \sim \langle {\cal H} \rangle$, the minimum $F (\mathbf{r})$ corresponds to the uniform magnetization modulus, $m^{2}(\mathbf{r}) = \bar{m}^2 \equiv- \tfrac{A_{2} }{2A_{4} }$. It allows one to return to our above analysis and to calculate the optimal configurations of the conical and skyrmion phases, Eq. \eqref{n-con}, \eqref{n-skyr}. In turn, the knowledge of $\mathbf{n} (\mathbf{r})$ determines the local energy density at the minimum, $E_{min}$, which is now a function of $m$ at each point on the plane.
If taking $E_{min}$ into account does not strongly affect the homogeneous solution, $ \bar{m}$, then we can limit ourselves by the first iteration of this procedure, i.e. 
\[  \left\{ m (\mathbf{r}) = \bar{m}\right\} \to   \left\{ \mathbf{n} (\mathbf{r}) \neq cst \right\}\to   \left\{ m (\mathbf{r}) \neq \bar{m} \right\}\,.\]  

Let us now provide more details to this consideration. For our purposes, we can write
$ E _{min} \simeq E_{0} m^{2} + E_{1} m + E_{2} \,.$
The $\mathbf{r}$-dependent coefficients $E_{i}$ can be determined by noting that ${\cal H} / m^{2}$ depends only on the ratio $b/m= H/H_{c2}$. Therefore, having determined the dependence of $E _{min}$ at a given point on $b/m$, we thereby find its dependence on $m$.

Introducing the notation
\begin{equation}
\begin{aligned}
\epsilon_0 & =\bar{m}^2\tfrac {D^2}{J}  \leftrightarrow m Ak_{0}^{2}\,, \quad 
\bar{b} =b  / \bar{m} \leftrightarrow  H /H_{c2} \,,
\\ 
\kappa & =\epsilon_0 /(8 A_{4} \bar{m}^4) = J k_0^2 / 4|A_2|\,, \quad 
\zeta  = m/ \bar{m}
  \,, 
 \end{aligned}
\label{defs}
\end{equation}
we can represent the free energy density in the form 
\begin{equation} 
\begin{aligned} 
F (\mathbf{r})&= \epsilon_0 \left( c_{0} \zeta^2+  c_{1}\zeta \bar{b} +  c_{2}\bar{b}^{2} +\tfrac {1}{8\kappa}  (\zeta^2-1)^2  \right) \,.
\end{aligned}
\end{equation}

Near $ T_c$ we have the estimate
\[\kappa \sim k_{0}^{2}/ \bar{m}^2 \sim k_{0}^{2}/ |\tau| \,,\]
therefore, at a sufficient distance from the transition point, $ k_{0}^{2} \ll |\tau| $, we obtain $\kappa \ll1$.
In this case, the minimum of $F (\mathbf{r})$ is achieved at $|\zeta|= 1$, and we return to the  above formulas, showing the energy gain of the conical helix phase.

As the temperature approaches the transition point from below, the minimum of $F (\mathbf{r})$ is realized at
\[ \zeta_0 = 1 - \kappa ( 2c_{0} + c_{1 }\bar{b}) + \mathcal{O} (\kappa^2) \,. \]
At this point the corresponding correction to the energy takes the form
\begin{equation}
\delta E  = - \tfrac{1}{2}  \epsilon_0 \kappa \,\langle ( 2c_{0} + c_{1 }\bar{b})^2 \rangle + \mathcal{O} (\epsilon_0  \kappa^2)  \,,  
\label{deltaE}
\end{equation}
here $\langle \ldots \rangle$ means averaging over $\mathbf{r} $. By restricting ourselves in Eq. \eqref{deltaE} to the first order in $\kappa$, we remain within the framework of the self-consistent approximation. Indeed, an attempt to calculate the next terms of the expansion leads to the necessity to take into account the term $(\tilde {\nabla} m )^2$ in \eqref{HamGL}, which, as follows from the expression for $\zeta_0$, is  proportional to $\kappa^2$.

According to \eqref{eq3}, we have  in the conical phase coefficients $c_{0}=c_{2} = -1/2$, $c_{1}
=0$ which are independent of $\mathbf{r} $, therefore $\delta E = - \tfrac{1}{2} \epsilon_0 \kappa$. In the skyrmion phase, the numerically found value $(2c_{0} + c_{1 }\bar{b})$ strongly depends on $\mathbf{r} $.
We note that the average value of $\langle \zeta_0 \rangle$ almost coincides in both phases. However, the noticeable dispersion of $\zeta_0 $ in the skyrmion phase over the skyrmion unit cell leads to a larger value of $|\delta E |$.
The results of calculating $\delta E$ in our model can be presented as follows.

Taking into account the correction $\delta E \sim \epsilon_0 \kappa $, the skyrmion phase becomes more favorable than the conical spiral as the temperature increases ($\kappa$ increases) in a certain region of the field.

\begin{figure}[h]
	\centering
 	\includegraphics[width=0.9\columnwidth]{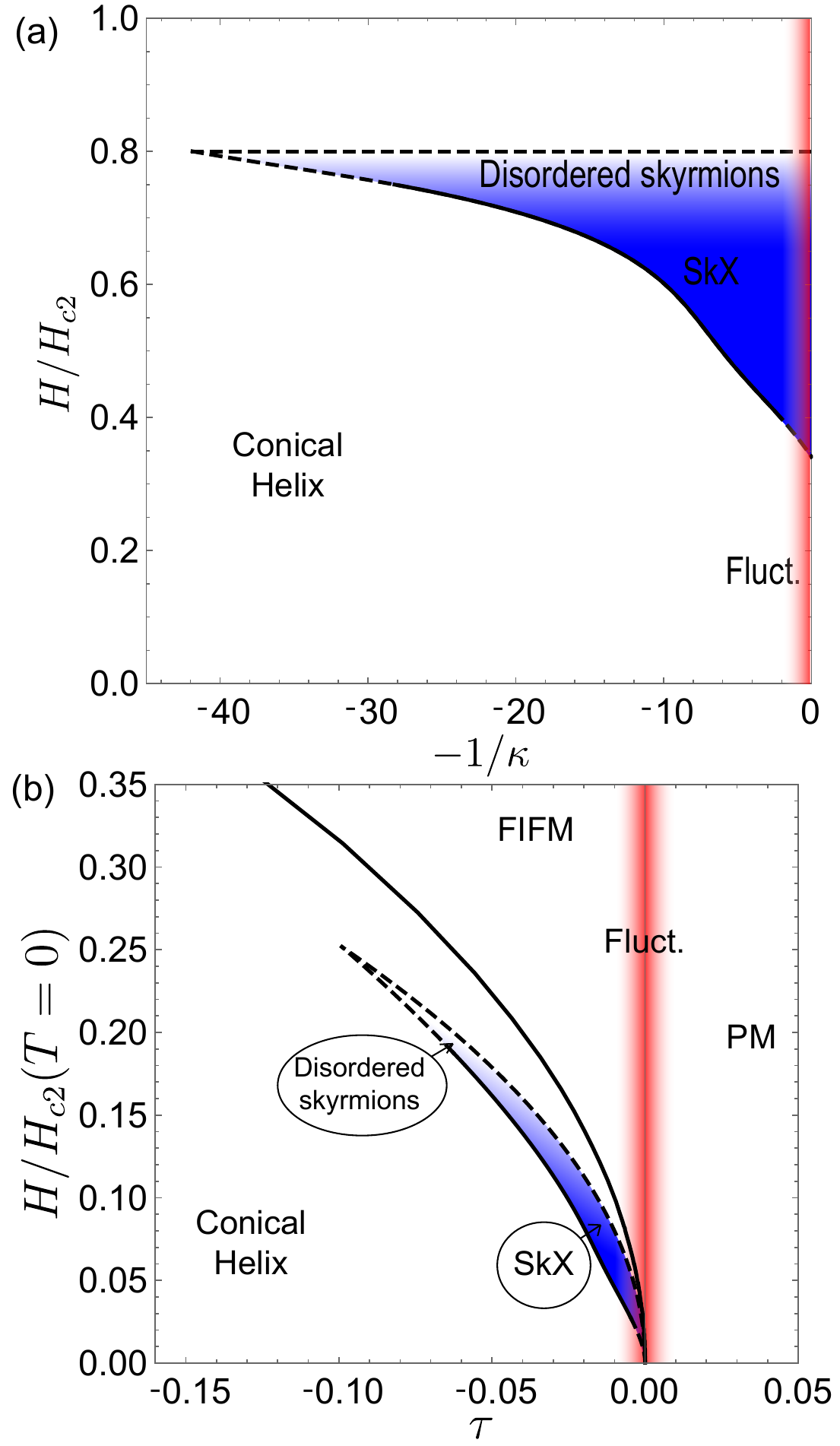}\\
	\caption{ Phase diagram near $T_{c}$, (a) the region of the skyrmion phase is shown as a function of the universal parameter $\kappa$ of proximity to the transition temperature, (b) $T-H$ diagram, temperature-field. } 
	\label{figSk}
\end{figure}

We saw above in Fig. \ref{fig1} that without taking into account the inhomogeneity of the magnetization modulus, the energy difference $\Delta E = (E_{sk} - E_{con}) $ in the entire range of fields was $ \Delta E = (0.7 \div 1.5) \cdot 10^{-2} \epsilon_0$. Since the coefficients $c_j$ are of the order of unity, the value of $ \delta E$ becomes comparable to $ \Delta E$ at $\kappa \sim 10^{-2}$.
The first time this happens is at $\kappa \simeq 0.04$ and at the maximum possible field value $H\simeq 0.8 H_{c2}$.
With the further increase, $ 0.04 < \kappa \alt 0.5$,  the skyrmion phase becomes lower in energy in the expanded region of field values, $ 0.4 \alt H/H_{c2} \alt 0.8$, as shown in Fig.\ \ref{figSk}. 
As $ \kappa$ increases further, our estimates obtained in the leading order in $ \kappa$ lose their applicability, and further terms must be taken into account, including the term $ ( { \nabla } m )^2$ in \eqref{HamGL}. Here we will not consider this region of strong fluctuations, $ \kappa \sim 1$.

Let us briefly discuss the phase diagram obtained in Fig. \ref{figSk}. The temperature range in which the skyrmion phase is realized is given by the universal inequality \[ \kappa \agt 0.04 \,,\] which does not contain specific values of $k_{0}$, $T_{c}$, and the magnitude of the localized moment, $S$. Assuming $S\sim 1$, we see that it corresponds to temperatures
\begin{equation}  
|\tau| = \frac{|T -T_{c}|}{T_{c}} \alt 25\, \frac{Jk_{0}^{2}}{T_c}  . 
\end{equation}
Under the condition $k_{0}=D/J \ll1$, this inequality does not contradict the assumption $|\tau| \ll 1$.
It should also be noted that the skyrmion phase boundary determined by the maximum field $H\simeq 0.8 H_{c2}$ corresponds to the melting of the skyrmion crystal \cite{Huang2020} and the disappearance of the topological charge density according to the law $\rho_{top} \sim 1/\log^{2}|H-0.8 H_{c2}|$, \cite{Timofeev2021} i.e. the phase transition between the first and second order. In this case, the average distance between skyrmions grows, and instead of a homogeneous magnetic phase, one can expect a mixture between the conical and skyrmion phases. For this reason, we highlighted the region of fields $H/ H_{c2}\in (0.6,0.8)$ in Fig. \ref{figSk} by the gradient background.

{\it Conclusion.}

We discuss the stability boundaries of the skyrmion lattice in non-centrosymmetric ferromagnets with the Dzyaloshinskii-Moriya interaction in external magnetic field. At low temperatures, the classical energy of the skyrmion lattice is slightly higher than the energy of the conical structure. We further show that careful consideration of dipole forces does not change this relation, despite some gain in the energy of the skyrmion lattice.
However, the possible inhomogeneity of the absolute value of the local magnetization as the temperature approaches $T_{c}$ leads to the gain in the free energy of the skyrmion lattice over the energy of the conical spiral. { This occurs already in the first order in the small parameter, $\sim J k_0^2/|T-T_c|$, at the level of the mean field theory, so the energy gain occurs even before proper consideration of the fluctuation corrections.} First of all, this is due to the fact that the local energy density of the skyrmion lattice is strongly inhomogeneous, and the energy correction is determined by the mathematical dispersion of this quantity. As a result, the  skyrmion lattice is realized at intermediate values of the magnetic field in a narrow temperature range near $T_{c}$. We find an estimate for the stability region of the skyrmion phase in terms of the model parameters.

{\bf Acknowledgments}.
The work of D.A. and V.T. was supported by the Russian Science Foundation, Grant No. 20-12-00147-$\Pi$. The work of D.A. and V.T. was partially supported by the Foundation for the Advancement of Theoretical Physics BASIS.


\begin{references}
	

\bibitem{Ishimoto1} K. Ishimoto, H. Yamaguchi, Y. Yamaguchi, J. Suzuki, M. Arai, M. Furusaka, Y. Endoh, J.Magn.Magn.Mat. {\bf 90\&91} (1990) 163.

\bibitem{Ishimoto2} K. Ishimoto, Y. Yamaguchi, J. Suzuki, M. Arai, M. Furusaka, Y. Endoh, Physica B {\bf 213\&214} (1995) 381.

\bibitem{Lebech_95} B. Lebech, P. Harris, J. Skov Pedersen, K. Mortensen, C. I. Gregory, N.R. Bernhoeft, M. Jermy, and S.A. Brown, J. Magn. Magn. Mater. \textbf{140}, 119 (1995).

\bibitem{Grigoriev2006} S.V. Grigoriev, S.V. Maleyev, A.I. Okorokov, Yu. O. Chetverikov, H. Eckerlebe, Phys.Rev. B, {\bf 73} (2006) 224440. 

\bibitem{Grigoriev2007_1}{S. V. Grigoriev, S. V. Maleyev, V. A. Dyadkin, D. Menzel, J. Schoenes, and H. Eckerlebe, Phys. Rev. B \textbf{76} 092407 (2007).}

\bibitem{Grigoriev2007_2}{S. V. Grigoriev, V. A. Dyadkin, Yu. O. Chetverikov, D. Menzel, J. Schoenes, A. I. Okorokov, H. Eckerlebe and S. V. Maleyev, Phys. Rev. B \textbf{76} 224424 (2007).}

\bibitem{MuhlbauerSience09_v323} {S. M\"uhlbauer, B. Binz, F. Jonietz, C. Pfleiderer, A. Rosch, A. Neubauer, R. Georgii, and P. Boni, Science {\bf 323} (2009) 915.}

\bibitem {Adams_PRL_2011} T. Adams, S. M\"uhlbauer, C. Pfleiderer, F. Jonietz, A. Bauer, A. Neubauer, R. Georgii, P. Boni, U. Keiderling, K. Everschor, M. Garst, and A. Rosch, Phys. Rev. Lett. {\bf 107} (2011) 217206. 

\bibitem{Grigoriev_Jetp_Lett_14} S. V. Grigoriev, N. Potapova, V. A. Dyadkin, E. V. Moskvin, Ch. Dewhurst,  S. V. Maleyev, JETP Letters {\bf 100} N. 3  (2014) 238-243.

\bibitem{PfleidererJPCM10_v22} C. Pfleiderer, T. Adams, A. Bauer, W. Biberacher, B. Binz, F. Birkelbach, P. Boni, C. Franz, R. Georgii, M. Janoschek, J. Phys.: Condens. Matter 22, 164207 (2010).

\bibitem{NeubauerPRL09_v102} A. Neubauer, C. Pfleiderer, B. Binz, A. Rosch, R. Ritz, P. G. Niklowitz, and P. B\"oni, Phys. Rev. Lett. 102,
186602 (2009).

\bibitem{Muenzer_PRB} W. Muenzer, A. Neubauer, T. Adams, S. M\"uhlbauer, C. Franz, F. Jonietz, R. Georgii, P. B\"oni, B. Pedersen, M. Schmidt, A. Rosch, and C. Pfleiderer, Phys. Rev. B {\bf 81}(2010) 041203.  


\bibitem{Moskvin_PRL_2013} E. Moskvin, S. Grigoriev, V. Dyadkin, H. Eckerlebe, M. Baenitz, M. Schmidt, and H. Wilhelm, Phys. Rev. Lett. {\bf 110} (2013) 077207.

\bibitem{Adams_PRL_2012} T. Adams, A. Chacon, M. Wagner, A. Bauer, G. Brandl, B. Pedersen, H. Berger, P. Lemmens, and C. Pfleiderer, Phys. Rev. Lett. {\bf 108} (2012) 237204.

\bibitem{White_PRL_2012} J.S. White, I. Levatic, A.A. Omrani, N. Egetenmeyer, K. Prsa, I.Z. Zivkovic, J.L. Gavilano, J. Kohlbrecher, M. Bartkowiak, H. Berger, and H M Ronnow, J. Phys.: Condens. Matter {\bf 24} (2012) 432201

\bibitem{Tokunaga} Y. Tokunaga, X. Z. Yu,	J. S. White,	H. M. Ronnow,	D. Morikawa,	Y. Taguchi, Y. Tokura,  Nature Communications {\bf 6}  (2015) 7638.


\bibitem{Bak} P.Bak, M.H.Jensen, J.Phys.\textbf{C13}, L881 (1980).

\bibitem{Kataoka_1} O. Nakanishi, A. Yanase, A. Hasegawa, M. Kataoka, Solid State Commun. {\bf 35} (1980) 995.

\bibitem{Buhrandt2013} S. Buhrandt, L. Fritz,  Phys. Rev. B 88, 195137 (2013).

\bibitem{Butenko2010} A.B. Butenko, A.A. Leonov, U.K. Rößler, and A. N. Bogdanov, Physical Review B {\bf 82} (2010) 052403.


\bibitem{Rybakov} F. N. Rybakov, A. B. Borisov, A. N. Bogdanov, Physical Review B {\bf 87} (2013) 94424.


\bibitem{Wilson} M. N. Wilson, A. B. Butenko, A. N. Bogdanov, and T. L. Monchesky, Phys. Rev. B {\bf 89} (2014)094411. 

\bibitem{Rossler2010} Ulrich K. R\"oßler, Andrei A. Leonov, Alexei N. Bogdanov,  J. Phys.: Conf. Ser. {\bf 303} (2010) 012105

\bibitem{Laliena2018}
V. Laliena,  G. Albalate, J. Campo, 
Phys. Rev. B 98, 224407 (2018)


\bibitem{Janoschek2013}
M. Janoschek,  M. Garst, A. Bauer, P.  Krautscheid, R. Georgii, P. Böni,  and C. Pfleiderer, 
 Phys. Rev. B, {\bf  87}, 134407 (2013). 

\bibitem{Yu_2010} X. Z. Yu, Y. Onose, N. Kanazawa,  J. H. Park, J. H. Han, Y. Matsui, N. Nagaosa, Y. Tokura,  Nature {\bf 465} (2010) 901-904. 

\bibitem{Yu_2011} X. Z. Yu, N. Kanazawa, Y. Onose, K. Kimoto, W. Z. Zhang, S. Ishiwata, Y. Matsui, Y. Tokura,  Nature Materials {\bf 10} (2011) 106-109.


\bibitem{Tokura_2012}  A. Tonomura, X. Yu, K. Yanagisawa, T. Matsuda, Y. Onose, N. Kanazawa, H. S. Park, and Y. Tokura, Nano Lett. {\bf 12} (2012) 1673-1677.

\bibitem{Yu2015} X. Yu, A. Kikkawa, D. Morikawa, K.Shibata,Yu. Tokunaga,
Y. Taguchi, and Y. Tokura, Phys. Rev. B {\bf 91} (2015) 054411


\bibitem{Monchesky} S. A. Meynell, M. N. Wilson, J. C. Loudon, A. Spitzig, F. N. Rybakov, M. B. Johnson, T. L. Monchesky, Physical Review B {\bf 90} (2014) 224419 

\bibitem{Kiselev1} F.N. Rybakov, A.B. Borisov , S. Blugel , N.S. Kiselev, Physical Review Letters {\bf  115} (2015) 11720.


\bibitem{Du_Nano_Lett} Haifeng Du, John P. DeGrave, Fei Xue, Dong Liang, Wei Ning, Jiyong Yang, Mingliang Tian, Yuheng Zhang, and Song Jin,   Nano Lett. 2014, 14, 4, 2026–2032.



\bibitem{Timofeev2021}
  V. E. Timofeev,  A. O. Sorokin, D. N.  Aristov, 
 Phys. Rev. B,  {\bf 103}, 094402 (2021) . 


\bibitem{Timofeev2022}
  V. E. Timofeev,  D. N.  Aristov, 
 Phys. Rev. B,  {\bf 105}, 024422 (2022).


\bibitem{bauer_PRB} A. Bauer and C. Pfleiderer,  Phys. Rev. B 85,  214418  (2012). 

\bibitem{Maleyev06PRB} S.V. Maleyev, Phys. Rev. B {\bf 73}, 174402 (2006).

\bibitem{Chubova_JETP_2017} Н.М.Чубова, Е.В. Москвин, В.А. Дядькин, Ch. Dewhurst, С.В. Малеев, С.В. Григорьев , ЖЭТФ 152, No. 5 (2017) стр. 933.

\bibitem{Kizhe} S. V. Grigoriev, A. S. Sukhanov, E. V. Altynbaev, S.-A. Siegfried, A. Heinemann, P. Kizhe, and S. V. Maleyev,  Phys. Rev. B {\bf 92} 220415(R) (2015).

 \bibitem{Grigoriev2011}  S.V. Grigoriev, N.A. Grigoryeva, K.S. Napol’skii,  et al. 
JETP Lett. 94, 635–641 (2011). 


\bibitem{Bauer2016}
A. Bauer, C.  Pfleiderer, in  Seidel, J. (Ed.)
{\it Topological Structures in Ferroic Materials: Domain Walls, Vortices and Skyrmions}, 
 Springer International Publishing, 
(2016)  

\bibitem{Utesov2017} O. I. Utesov and A. V. Syromyatnikov, Phys. Rev. B 95, 214420 (2017) 

\bibitem{Huang2020}
 P. Huang, T. Schönenberger,  M. Cantoni, L. Heinen,  A. Magrez,  A. Rosch,   F. Carbone,  H. M. R{o}nnow, 
 Nature Nanotechnology, {\bf 15}, 761  (2020)


 

\end{references}
\end{document}